\documentclass[3p,twocolumn,reversenotenum]{elsarticle}
\biboptions{sort&compress}
\usepackage[utf8]{inputenc}
\usepackage[english]{babel}

\usepackage{physics}
\usepackage{graphicx}
\usepackage{xcolor}
\usepackage{outlines}
\usepackage{romannum}
\usepackage{tensor}
\usepackage{float}
\usepackage{caption}
\usepackage{subcaption}
\usepackage{amsfonts}
\usepackage{amssymb}
\usepackage{physics}
\usepackage{graphicx}
\usepackage{xcolor}
\usepackage{MnSymbol}
\usepackage{float}
\usepackage{relsize}
\usepackage{romannum}
\usepackage{multicol}
\usepackage[colorlinks,citecolor=red,urlcolor=blue,bookmarks=false,hypertexnames=true
]{hyperref}
\newenvironment{equations}{\equation\aligned}{\endaligned\endequation}  
\newcommand{\beq}{\begin{equations}}
\newcommand{\eeq}{\end{equations}}
\newcommand{\boe}{\begin{outline}[enumerate]}
\newcommand{\eo}{\end{outline}}
\def\l{\left(}
\def\r{\right)}

\def\la{\mathcal{L}}

\def\={&=}
\def\p{\partial}

\def\-{\item[-]}
\def\G{G(u,m^2)}
\def\f21{\tensor[_2]{F}{_1}}
\def\m0{m_{0}}
\def\la0{\lambda_{0}}

\journal{Physics Letters B}

\begin{document}
\pagenumbering{arabic}
\begin{frontmatter}

\title{On the Observer Dependence of the Quantum Effective Potential}

\author[1]{Pallab Basu}
\ead{pallab.basu@wits.ac.za}
 \author[2]{S R Haridev\corref{cor1}}
 \ead{p20180460@hyderabad.bits-pilani.ac.in}
\author[2]{and Prasant Samantray} 
\ead{prasant.samantray@hyderabad.bits-pilani.ac.in}

\cortext[cor1]{Corresponding author}

\affiliation[1]{organization={Mandelstam Institute for Theoretical Physics, University of Witwatersrand},addressline={Johannesburg}, country={South Africa}}
\affiliation[2]{organization={BITS- Pilani Hyderabad Campus},
addressline={Jawahar Nagar, Shamirpet Mandal}, city={Secunderabad}, postcode={500078},
country={India}}

\begin{abstract}
In this short paper, we investigate the consequences of observer dependence of the quantum effective potential for an interacting field theory. Specializing to $d+2$ dimensional Euclidean Rindler space, we develop the formalism to calculate the effective potential. While the free energy diverges due to the presence of the Rindler horizon, the effective potential, which is a local function of space, is finite after the necessary renormalization procedure. We apply the results of our formalism to understand the restoration of spontaneously broken $\mathbb{Z}_2$ symmetry in three and four dimensions.
\end{abstract}

\begin{keyword}
Spontaneous symmetry breaking \sep Symmetry restoration \sep Accelerated observer 
\end{keyword}

\end{frontmatter}

\section{Motivation}
Ever since the discovery of Hawking's seminal result on black hole radiation \cite{Hawking1975}, there has been a tremendous amount of research in trying to understand the quantum origins of the horizon entropy. While a significant amount of effort has been devoted to the calculation of black hole entropy using the microscopics of the quantum field, the central object of interest has been the calculation of the quantum effective action in the Euclidean version of the relevant spacetime. This calculation is made even more tractable if one instead considers the spacetime as perceived by a uniformly accelerating observer in flat space, i.e. Rindler space \cite{doi:10.1119/1.1972547, PhysRevD.14.870}. Rindler space has an observer-dependent event horizon, and it essentially captures all the salient features of a real black hole spacetime without the accompanying mathematical complexity. While most of the work \cite{PhysRevD.51.5716, KABAT1995281,PhysRevD.50.2700, CALLAN199455,JSDowker} and recently \cite{PhysRevD.105.105003} have focused on the calculation of effective action, certain aspects of the quantum field theory are more ideally captured by the structure of the quantum effective potential instead. This has been attempted by \cite{TPadmanabhan_1983, https://doi.org/10.48550/arxiv.1703.05675, Castorina_2012} previously. However, these approaches are not general and fail to capture the correct behavior in dimensions other than the usual four. We, in this work, aim to remedy this issue and unravel certain subtleties related to the calculation of the effective potential in Euclidean Rindler space. In the later part of our paper, we apply our results to understand the observer dependence of the effective potential to better understand the general covariance of the phenomenon of spontaneous symmetry breaking in arbitrary dimensions. 
\section{Effective Potential in Rindler Space : The Set Up}\label{sec1}
The effective potential is the essential mathematical tool to understand different quantum effects related to the vacuum state of a quantum field theory \cite{PhysRevD.7.1888,colemanbook,zee}.  In this section, we calculate the effective potential in the Rindler frame for a free massive scalar field with the Euclidean action
\beq
S = \int dv_{x}\l\frac{1}{2}g^{\mu\nu}\p_{\mu}\phi\p_{\nu}\phi +\frac{m_{0}^{2}}{2}\phi^{2}\r,
\eeq
where $dv_{x} = d^{d+2}x\sqrt{g}$ is the $d+2$ dimensional invariant measure. Using saddle point approximation, one can integrate over the quantum fluctuations around classical field configuration leading to the effective action of the theory \cite{colemanbook,zee,Padmanabhan2016}. The corresponding Euclidean effective potential is 

\beq\label{vefff}
V_{eff} = \frac{1}{2}\m0^{2}\phi^{2}+ \frac{1}{2}\log\l -\nabla^{2}+ m_{0}^{2}\r,
\eeq
where $\nabla^{2}$ is the Laplacian  operator.  We can rewrite the last expression in Eq. \ref{vefff} using the coincidence limit of the Euclidean propagator $\G$ as \citep{PhysRevD.12.965,PhysRevD.105.105003,ooguri,BURGESS1985137} 
\beq\label{vefff1}
V_{eff} \= \frac{1}{2}\m0^{2}\phi^{2} +\frac{1}{2}\int_{0}^{m_{0}^{2}}dm^{2} \lim_{u\rightarrow 0} \G\\
\= \frac{1}{2}\m0^{2}\phi^{2}+V_{1},
\eeq
where $u$ is the invariant length element,  and in the second line of Eq \ref{vefff1}, we use $V_{1}$ as a shorthand for the integral.  The coincidence limit of the propagator naturally leads to ultraviolet divergence in the effective potential, which has to be renormalized.   
\subsection{A Tale of Two Observers : The Prelude}
We consider this theory in the frame of an accelerated observer using the Euclidean Rindler metric in $d+2$ dimensional space time given as
\beq\label{rindlermetric}
ds^{2} = \xi^{2}d\tau^{2}+d\xi^{2} + (dx^{1})^{2}+...+(dx^{d})^{2}.
\eeq
The reader should notice that we have not explicitly introduced any `acceleration parameter' usually denoted by `$g$' or `$\kappa$'. This is so since such a parameter can be absorbed into the Euclidean Rindler time, making it dimensionless. The only relevant quantity in this context is the inverse invariant horizon temperature given by $2\pi$. The metric in Eq. \ref{rindlermetric} is nothing but the usual Euclidean space expressed in polar coordinates. An inertial observer in these coordinates can be described as having coordinates $\xi=\xi_0$ and $\tau = \tau_0$. A uniformly accelerating/Rindler observer in the same coordinate system can be described as $\xi=\xi_0$ and $\tau$ not constant, but uniformly varying.

The non-trivial physics of Rindler space is due to the following observation. In Euclidean signature,  the space observed by an accelerated observer is topologically different from that of an inertial observer \cite{CHRISTENSEN197811}. The inertial observer sees the space with the usual topology $R^{d+2}$, but an accelerated observer sees the space with $R^{d+1}\times S^{1}$ topology, which is multiply connected. The difference in topology makes an accelerated observer perceive the vacuum of an inertial observer as thermal, which is the statement of the Fulling–Davies–Unruh effect \cite{PhysRevD.7.2850,PhysRevD.14.870,10.1143/PTP.88.1,doi:10.1142/S0217751X19410057}. In terms of Euclidean propagators,  one can define the propagator with respect to the vacuum of an inertial observer ($G_{2\pi}$) as well as with respect to the vacuum of an accelerated observer ($G_{\infty}$). Note that $G_{\infty}$ is defined on the covering space of the Euclideanized Minkoswki space after the removal of its origin \cite{JSDowker_1977,PhysRevD.18.1856,PhysRevD.36.3095,PhysRevD.36.3742,Fulling_2012}. The relationship between the two propagators is given by \cite{CHRISTENSEN197811,TROOST1979442,Fulling1987TemperaturePA}
\beq\label{g2pi}
G_{2\pi}\l \tau-\tau_{0}\r = \sum_{n=-\infty}^{\infty}G_{\infty}\l \tau-\tau_{0}+2\pi n\r.
\eeq
From Eq. \ref{g2pi} one can easily deduce that $G_{2\pi}$ is periodic in $\tau-\tau_{0}$ with a periodicity of $2\pi$. Comparing similar results in finite temperature field theory, one can interpret $G_{2\pi}$ as the thermalized version of $G_{\infty}$ \cite{CHRISTENSEN197811}.

So topologically, the problem of calculating effective potential in the Euclidean Rindler frame is similar to the calculation of the energy-momentum tensor in $R^{d+1}\times S^{1}$ (same as finite temperature field theory) space in $d+2$ spatial dimensions. This problem is addressed, and the solution is known in the literature \cite{PhysRevD.20.3052,birrell_davies_1982,PhysRevD.18.1856}. The renormalization procedure is as follows. One has to start with $G_{2\pi}$, the propagator with respect to the inertial vacuum, and integrate over the mass-squared parameter and subsequently take the coincident limit. However, from the perspective of the accelerating observer who perceives the space to be multiply-connected with topology $R^{d+1}\times S^{1}$, there are infinitely inequivalent ways to take the coincidence limit. The consistent renormalization procedure, which considers this subtle aspect, is implemented by subtracting the contribution arising out of $G_\infty$. Therefore, we calculate the renormalized one-loop effective potential by using $G_{\Delta}=G_{2\pi}-G_{\infty}$ instead. The physical condition we are imposing through this renormalization procedure is to set the vacuum energy with respect to the Rinlder vacuum to zero, as we should. Also, in finite temperature field theory, we renormalize the free energy by subtracting the zero temperature contribution \cite{Laine_2016}, comparing with this, the above renormalization is justified as $\lim_{a\rightarrow 0}G_{2\pi}=G_{\infty}$ \citep{CHRISTENSEN197811}. We would also like to emphasize the following point, which can be a potential source of confusion among readers. Note that $G_{2\pi}$ is the standard flat space propagator just expressed in polar coordinates \cite{rindlerprop}. So one can calculate the standard flat Euclidean space one-loop effective potential using the same $G_{2\pi}$. However, the renormalization must be performed such that the vacuum energy is set to zero with respect to the Euclidean vacuum. This is achieved by taking the coincidence limit in the usual space with $R^{d+1}$ topology and subsequently eliminating the singular contributions by including counterterms. In our work, as inertial and accelerated observers have different vacuum states, the renormalization of vacuum energy should be handled differently in a way that is consistent with their vacua. Therefore, to consistently study the physical effects from the perspective of an accelerated frame in Euclidean signature, one has to calculate the operator expectation values with respect to $G_{\Delta} =G_{2\pi}-G_{\infty}$ instead of just $G_{2\pi}$.

Also, $G_{2\pi}$ and $G_{\infty}$ have both finite and infinite parts. The infinite part of $G_{2\pi}$ matches with that of $G_{\infty}$ and gets canceled beautifully in the expression for $G_{\Delta}$. The crucial thing is that $G_{\Delta}$ also contains non-trivial subtraction of the finite parts, which gives the correct finite contribution to the effective potential (more details are given in \cite{PhysRevD.18.1856}). As a consistency check, the derived effective potential from $G_{\Delta}$ also gives the correct free energy and Rindler entropy density. Viewed this way, we conclude that the effects due to the presence of an event horizon as perceived by the accelerating observer are captured by following this consistent protocol for renormalization.

Unlike the effective action, the effective potential is a local function of space and therefore is finite after renormalization. Turning now to the formal expressions for $G_{2\pi}$ and $G_{\infty}$, one can always calculate these propagators using the Rindler \cite{RevModPhys.80.787} or from optical metric modes \cite{PhysRevD.51.5716,PhysRevD.52.3529}. For a $d+2$ dimensional Euclidean Rindler space with the metric Eq. \ref{rindlermetric},  $G_{\Delta}$ in the coincident limit is given by
 \cite{linet}
\beq\label{gdelta}
G_{\Delta} = \int \frac{d^{d}k}{\l 2\pi\r^{d}}\int \frac{d\nu}{\pi^{2}}e^{-\pi\nu} K^{2}_{i\nu}\l \xi \mu_{k}\r,
\eeq
where $\mu_{k}=\sqrt{k^{2}+m^{2}}$ and $K_{i\nu}(z)$ is the modified Bessel function of second kind.  One can calculate the total free energy (or the effective action) using Eq. \ref{gdelta} \cite{PhysRevD.105.105003}. Since this free energy gives the correct entropy for the Rindler space \cite{KABAT1995281}, we can expect to arrive at the correct effective potential using Eq. \ref{gdelta} in Eq. \ref{vefff}.

One can do the `$\nu$' and `$k$' integral in Eq. \ref{gdelta} by separating the `$\nu$' dependence in $K_{i\nu}(z)$ using Eq. 2.1 in \cite{doi:10.1080/10652469.2020.1737529} and the integral identity (p.693, Eq.6.596,3 in \cite{gradshteyn2007}). The resulting integral can be expressed in terms of elementary functions using (p.917, Eq. 8.432,6 in \cite{gradshteyn2007}) as
\beq\label{gdelta2}
G_{\Delta}=  \frac{1}{2\pi\l 4\pi\r^{\frac{d}{2}}\xi^{d}}&\int_{0}^{\infty}\frac{du}{\pi^{2}+u^{2}}\times\\
&\int_{0}^{\infty}\frac{ds}{s^{\frac{d}{2}+1}}e^{-\xi^{2} m^{2}s}e^{\frac{-\cosh(u/2)^{2}}{s}},
\eeq
where we introduced renormalized mass $m$.  One can calculate one-loop effective potential using Eq. \ref{gdelta2} in Eq. \ref{vefff1} as
\beq\label{v1}
V_{1}= \frac{-1}{\l 4\pi\r^{\frac{d}{2}+1}\xi^{d+2}}&\int_{0}^{\infty}\frac{du}{\pi^{2}+u^{2}}\times\\
&\int_{0}^{\infty}\frac{ds}{s^{\frac{d}{2}+2}}e^{-\alpha^{2}s}e^{\frac{-\cosh(u/2)^{2}}{s}},
\eeq
where $\alpha = m\xi$, and we subtract out the field independent term coming from $\lim_{\alpha\rightarrow 0}V_{1}$ term. It turns out that there is no closed-form expression for the complete integral in Eq.\ref{v1}. However, one can derive an approximate result given as (see \ref{appendix2} for more details) 
\beq\label{main}
V_{1}\approx -\frac{2}{\l 4\pi\r^{\frac{d+2}{2}}\pi^{\frac{3}{2}}}\frac{\alpha^{\frac{d+1}{2}}}{\xi^{d+2}}K_{\frac{d+1}{2}}\l 2\alpha\r.
\eeq
This approximation is particularly effective for limiting values of $\alpha$ and gives accurate qualitative behavior and using this result, we can study the behavior of effective potential for observers who are moving with both high and low acceleration values. Also, we can check the validity of our approximation by comparing these results with the field theory results at finite temperatures (see \ref{app1}).
\subsection{Observers with high values of acceleration for $d=1,2$}\label{subsec1}
Near horizon observers are the observers with high acceleration. We can study this limit by taking  $\alpha\rightarrow 0$ in Eq. \ref{main}. This limit is equivalent to the high-temperature limit in finite temperature field theory. It may be tempting to perform a Taylor series expansion around $\alpha =0$ in Eq. \ref{v1}. However, as a function of $\alpha$, $V_{1}$ is not analytic at $\alpha =0$ as  the '$s$' integral diverges for $n^{th}$ Taylor term when $n\geq d/2+1$. Nevertheless, it still gives the correct leading order term for the four-dimensional case. For four dimensions, $d=2$, the second term in the series expansion around $\alpha =0$ in Eq. \ref{v1}  gives (after restoring $m$ and $\xi$)
\beq\label{4drindleractual}
\lim_{\alpha\rightarrow 0}V_{1} = \frac{m^{2}}{96\pi^{2}\xi^{2}}.
\eeq
One can now take the interpretation that an observer with the trajectory given by constant $\xi$ perceives an Unruh temperature of $\xi = 1/2\pi T$. We immediately observe that Eq. \ref{4drindleractual} agrees with the field theory results at finite temperature (see Eq.\ref{appv4hight}). However, this procedure fails for other dimensions. So for general dimensions,  one can study the near horizon limit using Eq. \ref{main} as a good approximation for the effective potential. For three dimensions, taking $d=1$ and ($\alpha\rightarrow 0$) in Eq. \ref{main} gives
\beq\label{3drindler}
\lim_{\alpha\rightarrow 0}V_{1}\approx \frac{m^{2}}{8\pi^{3}\xi}\l 1-\gamma_{E}\r -\frac{m^{2}}{8\pi^{3}\xi}\log\l m^{2}\xi^{2}\r,
\eeq
where $\gamma_{E}$ is Euler gamma and we remove the terms independent of $m$. Considering the Unruh temperature $\xi=1/2\pi T$, one can compare Eq. \ref{3drindler} with that of finite temperature results ( see Eq. \ref{fthkv}). The temperature dependence of the effective potential perfectly matches that of finite temperature results (see Eq. \ref{fthkv}), with only the coefficients not agreeing perfectly at this order of approximation. This can of course be improved by considering higher-order corrections. Similar calculations in four dimensions, i.e, $d=2$ and $\alpha\rightarrow 0$ in Eq. \ref{main} gives 
\beq\label{4dapprox}
\lim_{\alpha\rightarrow 0}V_{1}\approx \frac{m^{2}}{16\pi^{3}\xi^{2}}.
\eeq
Which is a sufficiently good approximation of Eq. \ref{4drindleractual}. Eq. \ref{main} can in fact be used to study the behavior of effective potential in Euclidean Rindler space in general dimensions.
\subsection{Observers with low values of acceleration for $d=1,2$}
Now we consider the other limit, the observer who accelerates slowly and stays far from the horizon ($\alpha\rightarrow \infty$ in Eq. \ref{main}). In three dimensions, taking $d=1$ and $\alpha\rightarrow\infty$ in Eq. \ref{main} gives
\beq\label{low3d}
\lim_{\alpha\rightarrow \infty}V_{1} \approx -\frac{e^{-2\alpha}\alpha^{\frac{1}{2}}}{8\pi^{\frac{5}{2}}\xi^{3}}.
\eeq
One can compare this result with the finite temperature results in Eq. \ref{applow3d} by taking the Unruh temperature ($\xi = 1/2\pi T$). The leading exponential decay behavior is the same as that of finite temperature. However, polynomial powers of the temperature change by a factor of $1/2$. This may be because of the approximation scheme we are using. In the case of four dimensions, similar limit yields \beq\label{low4d}
\lim_{\alpha\rightarrow \infty}V_{1} \approx -\frac{e^{-2\alpha}\alpha}{16\pi^{3}\xi^{4}}.
\eeq
Therefore we conclude that for a slowly accelerating observer with proper acceleration `$a$', the corrections to effective potential are exponentially suppressed by the Boltzmann factor $e^{-m/a}$. Since, exponential suppression is the leading order behavior of the modified Bessel function ($\lim_{z\rightarrow\infty}K_{\nu}(z)\sim e^{-z}$). So from Eq. \ref{main}, one can expect this behavior in arbitrary dimensions as well.
\section{Symmetry Restoration in $\lambda\phi^{4}$ Theory}\label{sec2}
We now apply our results to investigate the special case of understanding whether the breakdown of discrete $\mathbb{Z}_2$ symmetry is restored for a certain class of observers. It is standard procedure to study sponataneous symmetry breaking using effective potential \cite{colemanbook,zee}. Armed with the results from the previous section (section \ref{sec1}),  we can now easily understand spontaneous symmetry breaking in the Rindler frame. As the Rindler frame is another coordinate system for the standard flat space, comparing it with the standard flat space results can help to understand the frame dependence of spontaneous symmetry breaking. So we  consider a self-interacting massive scalar field theory with the action
\beq
S = \int dv_{x}\;  \l\frac{1}{2}g^{\mu\nu}\p_{\mu}\phi\p_{\nu}\phi +\frac{m_{0}^{2}}{2}\phi^{2}+\frac{\lambda_{0}}{4!}\phi^{4}\r.
\eeq
The corresponding one loop effective potential is 
\beq\label{vefff2}
V_{eff} = \frac{1}{2}\m0^{2}\phi^{2}+\frac{\lambda_{0}}{4!}\phi^{4}+ \frac{1}{2}\log\l -\nabla^{2}+ M^{2}\r,
\eeq
where $M^{2}=m_{0}^{2}+\lambda_{0}\phi^{2}/2$ and $\phi$ is a constant.  Note that the $V_{1}$ corresponding to this effective potential (Eq. \ref{vefff2}) differs from the $V_{1}$ in Eq. \ref{vefff1} only by the formal replacement $m_{0}^{2} \rightarrow M^{2}$. Therefore, for arbitrary dimensions we can take Eq. \ref{main} as the $V_{1}$ for the interacting field theory by replacing $m_{0}^{2}$ with $M^{2}$ as
\beq\label{main2}
V_{1}\approx -\frac{2}{\l 4\pi\r^{\frac{d+2}{2}}\pi^{\frac{3}{2}}}&\frac{\l m^{2}+\lambda\phi^{2}/2\r^{\frac{d+1}{4}}}{\xi^{\frac{d+3}{2}}}\times\\
&K_{\frac{d+1}{2}}\l 2\xi\sqrt{ m^{2}+\frac{\lambda}{2}\phi^{2}}\r,
\eeq
where $m$ and $\lambda$ are now the renormalized parameters. From Eq. \ref{main2}, we can see that the effective potential is explicitly dependent on $\xi$ or the trajectory of the accelerated observer \cite{https://doi.org/10.48550/arxiv.1703.05675}. So the re-normalized parameters of the theory, like mass, can also depend on the observer, which point towards the observer dependence of the spontaneous symmetry breaking in the system. Also, we can interpret the result for the effective potential in the following way - Consider discretizing the radial coordinate in the Euclidean Rindler frame. At the each radial lattice point say ``i" (with the usual rotational invariance) we have a different form of the effective potential dependent on the label ``i" - i.e. each observer fixed at such a lattice point ``i" perceives a different effective potential and therefore a different coefficient of $\phi^2$. It is well known in the literature that in standard flat space, the vacuum state configuration of a scalar field in $\lambda\phi^{4}$ interaction breaks the $Z_{2}$ symmetry for $m^{2}<0$ \cite{zee, colemanbook}. It is also known that at finite temperatures, this symmetry is restored above a certain critical temperature \cite{KIRZHNITS1972471,PhysRevD.9.3320,PhysRevD.9.3357}. By appealing to the thermalization theorem \cite{10.1143/PTP.88.1}, we therefore can expect similar restoration of broken symmetries in accelerated frames as shown in \cite{TPadmanabhan_1983,Castorina_2012}. Similar to the finite temperature calculations, we can calculate critical acceleration for the symmetry restoration by imposing \cite{PhysRevD.9.3320}
\beq\label{condition}
\frac{\p^{2} V_{eff}}{\p \phi^{2}}\Big|_{\phi=0}=0.
\eeq
In Eq. \ref{condition}, we defined the critical acceleration as the one which vanishes the re-normalized mass. Considering the effective potential as discretized along radial direction the condition (Eq. \ref{condition}) should be thought of as imposing a renormalization condition for absorbing the quadratic divergence at each such lattice point. Using this (Eq. \ref{condition}) in Eq. \ref{main2} with $m^{2}<0$ one can calculate the approximate critical acceleration in four dimension as
\beq\label{ac4}
a_{c}^{2}=\frac{16\pi^{3}}{\lambda}m^{2}e^{2m\xi},
\eeq
where we choose the trajectory $\xi = 1/a_{c}$.  This (Eq. \ref{ac4}) critical acceleration in the near horizon limit ($m\xi \rightarrow 0$) reproduces the results in \cite{TPadmanabhan_1983, Castorina_2012} within the limitations of our approximation.  It is important to draw attention to the fact that the methods used in \cite{Castorina_2012,TPadmanabhan_1983} do not generalize to arbitrary dimensions and actually yield incorrect results for dimensions other than four.  In \cite{Castorina_2012}, authors calculated the critical acceleration using Taylor expansion as discussed in subsection \ref{subsec1}, this procedure fails in the following two ways in dimensions other than four. First, it fails to obtain the correct functional dependence of the effective potential in the $m\xi\rightarrow 0$ limit. Second, in four dimensions, the divergence part of the effective potential naturally separates despite performing this Taylor expansion in $m\xi$ at the leading order, which doesn't happen in other dimensions; one needs to specify the renormalization procedure. One can easily understand this in three dimensions, the critical acceleration in three dimensions by taking $d=1$ in Eq. \ref{main2} and using Eq. \ref{condition} is
\beq\label{3dcriticac}
-\frac{m^{2}}{2}+\frac{\lambda a_{c}}{8\pi^{3}}K_{0}\l \frac{2m}{a_{c}}\r = 0.
\eeq
As $\lim_{z\rightarrow 0}K_{\nu}\l z\r\rightarrow +\infty$ we can conclude that the broken symmetry will be restored in three dimensions after a critical acceleration $a_{c}$. In the limit, $m\xi\rightarrow 0$ (near horizon limit) the leading order behaviour of Eq. \ref{3dcriticac} is
\beq
-\frac{m^{2}}{2}+\frac{\lambda a_{c}}{8\pi^{3}}\log\l\frac{a_{c}}{m}\r = 0.
\eeq
Now considering the Unruh temperature ($a_{c} =2\pi T_{c}$), one can reproduce the results of finite temperature field theory as in \cite{EINHORN1993611,Fujimoto1987}. Also, note that one can't obtain the logarithmic functional dependence on the critical acceleration using a Taylor expansion as in \cite{Castorina_2012}.

As in the case of finite temperature field theory \cite{PhysRevD.9.3320}, a precise calculation of critical acceleration requires higher loop corrections in the effective potential \cite{Castorina_2012}. But in this paper, we are interested in the observer dependence of the spontaneous symmetry breaking, for which the one-loop effective potential is a good approximation.

\section{Summary and Discussion}\label{sec5}
In this work, we calculated the one-loop effective potential in the Rindler space for arbitrary dimensions using the exact mode functions of the Rindler space. (Eqs \ref{main}, \ref{4drindleractual},\ref{4dapprox},\ref{3drindler},\ref{low3d} and \ref{low4d}) are the central results of our analyses.

In section \ref{sec1}, we discuss the proper renormalization procedure and derive a simple approximate result for the one-loop effective potential in arbitrary dimensions (Eq. \ref{main}). Using this result, we studied the nature of effective potential for a slowly accelerating and rapidly accelerating observer. Looking at Eq. \ref{main} or Eq. \ref{v1}, we can conclude that in the near horizon limit, the effective potential behaves $a^{d+2}$ for $d+2$ dimensional Rindler observer with acceleration $a$ at the leading order. The successive corrections for three and four dimensions are also studied using Eq. \ref{main}, and the results agree with the finite temperature results. For a slowly accelerating observer, the corrections to the effective potential due to acceleration are exponentially suppressed (Eq. \ref{low3d} and Eq. \ref{low4d}).

In section \ref{sec2}, we extend these studies to interacting field theory and understand symmetry restoration in accelerated frames. Using the effective potential, we studied symmetry restoration and compared the results with the field theory results in finite temperature. Similar calculations were done in literature with methods specific to four dimensions only \cite{Castorina_2012} or using the thermalization theorem \cite{TPadmanabhan_1983,PhysRevD.99.125018,https://doi.org/10.48550/arxiv.1703.05675}.

In contrast, we generalize the computation of effective potential for arbitrary dimensions for Euclidean Rindler space - which is our novel result. By applying to a specific interacting field theory with a broken $\mathbb{Z}_2$ symmetry, we conclude that the broken $Z_{2}$ symmetry in the standard flat space is restored for an accelerating observer after a critical acceleration $a_{c}$ in both three and four dimensions. Since our methods allow for similar analysis in arbitrary dimensions, we can conclude that the phenomenon of spontaneous symmetry breakdown is indeed observer dependent.
\section*{Acknowledgement}
We thank Prof. Dawood Kothawala for his useful comments. We also thank Prof. Steven Strogatz for his helpful suggestions for solving Eq. \ref{v1}. Research of P.S is partially supported by CSIR grant 03(1350)/16/EMR-II Govt. of India, and also by the OPERA fellowship from BITS-Pilani, Hyderabad Campus. H. R is supported by CSIR grant 03(1350)/16/EMR-II Govt. of India.

\appendix
\section{Finite Temperature Field Theory using Heat kernel}\label{app1}
In this appendix, we calculate the effective potential for finite temperature field theory using the heat kernel method. The derivation of these results using Green's functions has been done in \cite{Laine_2016}. The $d+2$ dimensional finite temperature quantum field theory in equilibrium is equivalent to a Euclidean field theory where the periodicity of imaginary time plays the role of inverse temperature. Therefore, one needs to study field theory in $S^{1}\times R^{d+1}$ manifold. For a massive scalar field, the heat kernel in the coincident limit in $S^{1}\times R^{d+1}$ manifold is given by
\beq
K(s,x,x) = \frac{1}{\l 4\pi s\r^{\frac{d+2}{2}}}\sum_{n=-\infty}^{\infty}e^{-\frac{n^{2}\beta^{2}}{4 s}}e^{-sm ^{2}}.
\eeq
In this form, one can see that $n=0$ corresponds to the standard flat space heat kernel. As we are interested in the finite temperature effect, we subtract out the flat space term to get
\beq
K(s,x,x) = \frac{2}{\l 4\pi s\r^{\frac{d+2}{2}}}\sum_{n=1}^{\infty}e^{-\frac{n^{2}\beta^{2}}{4 s}}e^{-sm ^{2}}.
\eeq
Corresponding one-loop effective potential is given by
\beq\label{ftvd}
V \= -\frac{1}{2}\int_{0}^{\infty}\frac{ds}{s}K(s,x,x)\\
\=-2\l\frac{m}{2\pi\beta}\r^{\frac{d+2}{2}}\sum_{n=1}^{\infty}\frac{1}{n^{\frac{d+2}{2}}}K_{\frac{d+2}{2}}\l m n \beta\r,
\eeq
where $K_{\nu}(z)$ is the modified Bessel function of second kind. One can get the same effective potential using method used in \cite{HARIDEV2022137489}. Using this result (Eq. \ref{ftvd}) one can study the high temperature and the low temperature expansion of the effective potential.  For three dimension, taking $d=1$ in Eq. \ref{ftvd}, the leading order high temperature expansion ($ m\beta\rightarrow 0$ in Eq. \ref{ftvd}) gives
 \beq\label{fthkv}
\lim_{m\beta\rightarrow 0} V \= - \frac{\zeta(3)}{2\pi\beta^{3}}+\frac{m^{2}}{8\pi\beta}-\frac{m^{2}}{4\pi\beta}\log\l m\beta\r.
 \eeq
This is in agreement with the results in \citep{EINHORN1993611,Fujimoto1987}. Similar calculations in four dimensions ($d=2$ in Eq. \ref{ftvd}) gives
 \beq\label{appv4hight}
\lim_{m\beta\rightarrow 0} V= -\frac{\pi^{2}}{90\beta^{4}}+\frac{m^{2}}{24\beta^{2}}.
 \eeq
Similarly,  we can study the low temperature expansion. For three dimension, with $d=1$ and $m\beta\rightarrow \infty$ in Eq. \ref{ftvd}, gives
 \beq\label{applow3d}
 \lim_{m\beta\rightarrow \infty}V = -\frac{m}{2\pi\beta^{2}}e^{-m\beta}.
 \eeq
 Similar expansions in four dimensions gives
 \beq\label{applow4d}
 \lim_{m\beta\rightarrow \infty}V = -\frac{m}{\l 2\pi\r^{\frac{3}{2}}\beta^{\frac{5}{2}}}e^{-m\beta}.
 \eeq
 These are in agreement with the results in \cite{Laine_2016}.

 \subsection{Mass correction in $\lambda\phi^{4}$ theory}
 Now if we are considering $\lambda\phi^{4}$ theory and interested in calculating the one-loop effective potential, one can just replace $m^{2}$ in Eq. \ref{fthkv} with $M^{2}= m^{2}+\frac{\lambda}{2} \phi^{2}$, where $m$  and $\lambda$ are the renormalized parameters of the scalar field and $\phi$ is the constant field configuration.  In three dimensions,  the total potential is
 \beq
 V = \frac{m^{2}}{2}\phi^{2}+\frac{\lambda}{4!}\phi^{4}+\frac{M^{2}T}{8\pi}-\frac{M^{2}T}{8\pi}\log\l \frac{M^{2}}{T^{2}}\r.
 \eeq
 The total mass of the system is defined as
 \beq
 \frac{\p^{2} V}{\p\phi^{2}}\Big|_{\phi\rightarrow 0} = \mu^{2}.
 \eeq
 This gives
 \beq
 \mu^{2}= m^{2}+\frac{\lambda T}{4\pi}\log\l \frac{T}{m}\r.
 \eeq
 This is in agreement with Eq. 3.4 of \citep{EINHORN1993611} and  Eq. 4 of \citep{Fujimoto1987}. One do similar calculations in four dimensions, comparing with \ref{appv4hight}
 \beq\label{appv4hightphi4}
 V=  \frac{m^{2}}{2}\phi^{2}+\frac{\lambda}{4!}\phi^{4}+\frac{\lambda}{24\beta^{2}}\phi^{2} .
 \eeq
 Similar to the three dimensional case, one can calculate the mass corrections in $\lambda\phi^{4}$ theory due to temperature as
 \beq
 \mu^{2}= m^{2}+\frac{\lambda T^{2}}{12}.
 \eeq
 
 \section{Approximation Scheme for Main Result}\label{appendix2}
In this appendix, we show that Eq. \ref{main} is indeed a good approximation to Eq. \ref{v1}.  We essentially draw upon the inspiration of applications of boundary layer theory to differential equations with a small/large parameter. Consider the integral
\beq\label{appv1}
V_{1}= \frac{-1}{\l 4\pi\r^{\frac{d}{2}+1}\xi^{d+2}}&\int_{0}^{\infty}\frac{du}{\pi^{2}+u^{2}}\times\\
&\int_{0}^{\infty}\frac{ds}{s^{\frac{d}{2}+2}}e^{-\alpha^{2}s}e^{\frac{-\cosh(u/2)^{2}}{s}}.
\eeq
One can use an approximation method by first scaling $s$ as $s=\alpha^{2}t$ and then splitting the integral into two parts as
\beq\label{vappr1}
V_{1}=  &\frac{-1}{\l 4\pi\r^{\frac{d}{2}+1}\xi^{d+2}\alpha^{d+2}}\times\\
&\Bigg(\int_{0}^{1/\alpha^{2}}\frac{dt}{t^{\frac{d}{2}+2}}e^{-\alpha^{4}t}\int_{0}^{\infty}\frac{du}{\pi^{2}+u^{2}}e^{\frac{-\cosh(u/2)^{2}}{\alpha^{2}t}}\\
&+\int_{1/\alpha^{2}}^{\infty}\frac{dt}{t^{\frac{d}{2}+2}}e^{-\alpha^{4}t}\int_{0}^{\infty}\frac{du}{\pi^{2}+u^{2}}e^{\frac{-\cosh(u/2)^{2}}{\alpha^{2}t}}\Bigg).
\eeq
In the first term in Eq. \ref{vappr1} $\alpha^{2}t<1$, so one can do the $u$ integral using Laplace method and the resulting $t$ integral in the leading order can be done in terms of Bessel function of the second kind.  The leading order approximation is
\beq\label{appmain}
V_{1}\approx -\frac{1}{\l 4\pi\r^{\frac{d+2}{2}}\pi^{\frac{3}{2}}}\frac{\alpha^{\frac{d+1}{2}}}{\xi^{d+2}}K_{\frac{d+1}{2}}\l 2\alpha\r.
\eeq
The correction from the second integral in Eq. \ref{vappr1} is of the order $O(\alpha^{4})$, so one can safely neglect that in comparison with the leading order term. In order to obtain the other extreme limit for the parameter $\alpha$, we now scale $s =t/\alpha^{2}$ in Eq. \ref{appv1} to get
\beq\label{vappr2}
V_{1}= & \frac{-\alpha^{d+2}}{\l 4\pi\r^{\frac{d}{2}+1}\xi^{d+2}}\times\\
&\Bigg(\int_{0}^{\alpha^{2}}\frac{dt}{t^{\frac{d}{2}+2}}e^{-t}\int_{0}^{\infty}\frac{du}{\pi^{2}+u^{2}}e^{\frac{-\alpha^{2}\cosh(u/2)^{2}}{t}}\\
&+\int_{\alpha^{2}}^{\infty}\frac{dt}{t^{\frac{d}{2}+2}}e^{-t}\int_{0}^{\infty}\frac{du}{\pi^{2}+u^{2}}e^{\frac{-\alpha^{2}\cosh(u/2)^{2}}{t}}\Bigg).
\eeq
Now as $\alpha\rightarrow\infty$, one can do the first integral in Eq. \ref{vappr2} using the Laplace method which gives the same result as Eq. \ref{appmain}. Therefore, for the two extreme conditions considered in this paper, Eq. \ref{main} is a good approximation to Eq. \ref{v1}.

\begin{figure}[H]
\begin{subfigure}{0.49\linewidth}
\centering
\includegraphics[width=\linewidth]{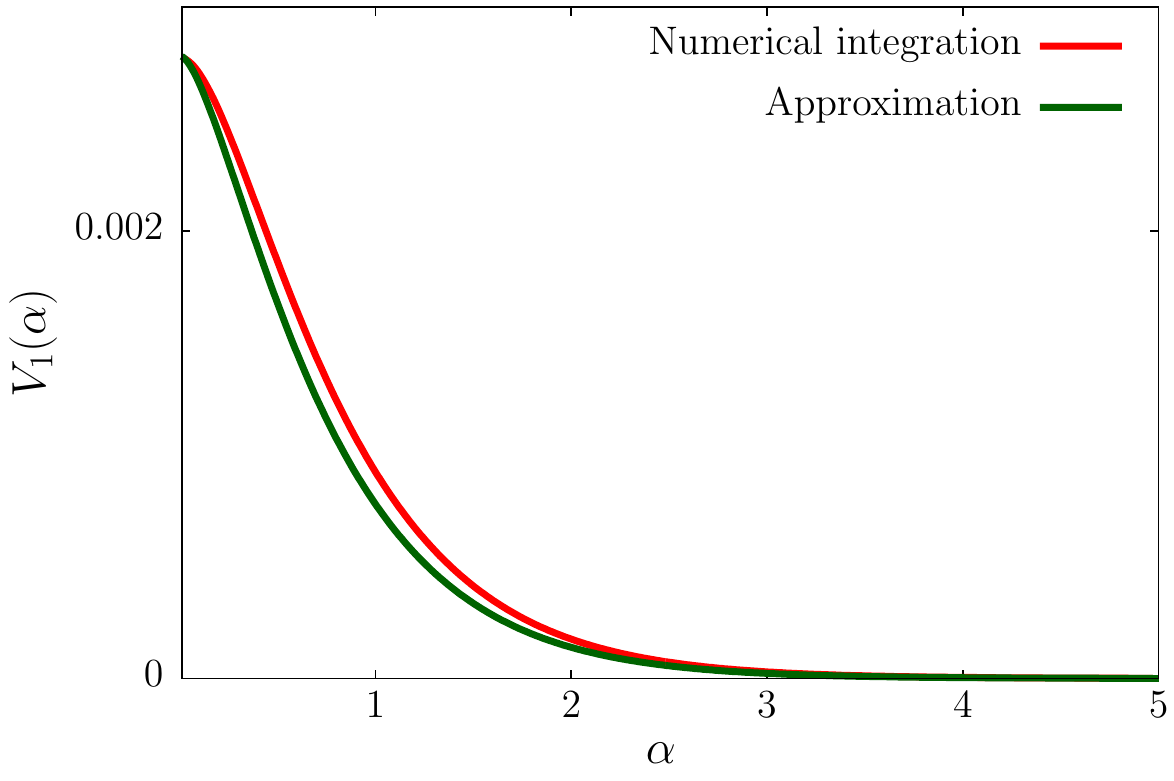}
\caption{Three dimensions}
\label{a}
\end{subfigure}\hfill
\begin{subfigure}{0.46\linewidth}
\centering
\includegraphics[width=\linewidth]{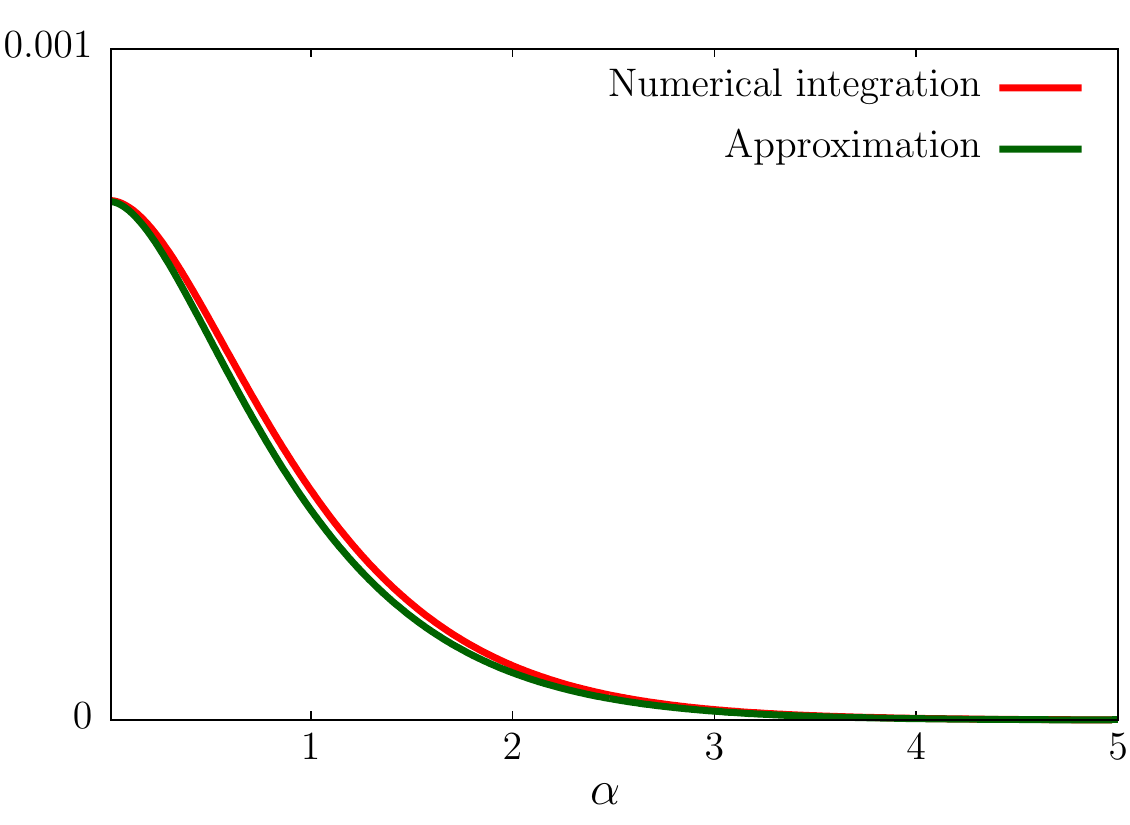}
\caption{Four dimensions}
\label{b}
\end{subfigure}\hfill
\captionsetup{justification=raggedright,singlelinecheck=false}
\caption{Plot compare the approximate result Eq. \ref{appmain} with the numerical integration result for Eq. \ref{appv1} in $d=1$ and $d=2$ cases .  Red plot shows the numerical integration result and the green plot shows the approximate result Eq. \ref{appmain} scaled with an appropriate constant.}
\label{fig1}
\end{figure}

In fact, one can compare the entire function (Eq. \ref{appv1}) to the approximation in Eq. \ref{appmain} by numerically integrating Eq. \ref{appv1} for specific values of $d$.  The comparison plot is given in Fig. \ref{fig1}. From figure Fig. \ref{fig1} one can conclude that up to a scaling Eq. \ref{appmain} is a good approximation for Eq. \ref{appv1}. Additionally, since this scaling depends on the numerical coefficients, in order to get a reasonably accurate qualitative insight we can always replace Eq. \ref{appv1} with Eq. \ref{appmain}.
        
\bibliographystyle{elsarticle-num-names} 
 \bibliography{reference}
\end{document}